\documentclass[preprint,12pt]{elsart}

\usepackage{amssymb}

\usepackage{lineno}
\usepackage{amssymb,latexsym,mathrsfs,amsmath,subfigure,epsfig,rotating,color,fancyheadings,syntonly,wrapfig,psfrag,wasysym,graphicx}

\journal{Astroparticle Physics}

\begin{document}

\begin{frontmatter}



\title{Measurement of the Group Velocity of Light in Sea Water at the ANTARES Site}


\author[UPV]{S.~Adri\'an-Mart\'inez},
\author[CPPM]{I. Al Samarai},
\author[Colmar]{A. Albert},
\author[UPC]{M.~Andr\'e},
\author[Genova]{M. Anghinolfi},
\author[Erlangen]{G. Anton},
\author[IRFU/SEDI]{S. Anvar},
\author[UPV]{M. Ardid},
\author[NIKHEF]{A.C. Assis Jesus},
\author[NIKHEF]{T.~Astraatmadja\thanksref{tag:1}},
% ...

\nopagebreak[3]
\address[UPV]{\scriptsize{Institut d'Investigaci\'o per a la Gesti\'o Integrada de les Zones Costaneres (IGIC) - Universitat Polit\`ecnica de Val\`encia. C/  Paranimf 1 , 46730 Gandia, Spain.}}\vspace*{0.15cm}
\nopagebreak[3]
\vspace*{-0.20\baselineskip}
\nopagebreak[3]
\address[CPPM]{\scriptsize{CPPM, Aix-Marseille Universit\'e, CNRS/IN2P3, Marseille, France}}\vspace*{0.15cm}
\nopagebreak[3]
\vspace*{-0.20\baselineskip}
\nopagebreak[3]
\address[Colmar]{\scriptsize{GRPHE - Institut universitaire de technologie de Colmar, 34 rue du Grillenbreit BP 50568 - 68008 Colmar, France }}\vspace*{0.15cm}
\nopagebreak[3]
\vspace*{-0.20\baselineskip}
\nopagebreak[3]
\address[UPC]{\scriptsize{Technical University of Catalonia, Laboratory of Applied Bioacoustics, Rambla Exposici\'o,08800 Vilanova i la Geltr\'u,Barcelona, Spain}}\vspace*{0.15cm}
\nopagebreak[3]
\vspace*{-0.20\baselineskip}
\nopagebreak[3]
\address[Genova]{\scriptsize{INFN - Sezione di Genova, Via Dodecaneso 33, 16146 Genova, Italy}}\vspace*{0.15cm}
\nopagebreak[3]
\vspace*{-0.20\baselineskip}
\nopagebreak[3]
\address[Erlangen]{\scriptsize{Friedrich-Alexander-Universit\"at Erlangen-N\"urnberg, Erlangen Centre for Astroparticle Physics, Erwin-Rommel-Str. 1, 91058 Erlangen, Germany}}\vspace*{0.15cm}
\nopagebreak[3]
\vspace*{-0.20\baselineskip}
\nopagebreak[3]
\address[IRFU/SEDI]{\scriptsize{Direction des Sciences de la Mati\`ere - Institut de recherche sur les lois fondamentales de l'Univers - Service d'Electronique des D\'etecteurs et d'Informatique, CEA Saclay, 91191 Gif-sur-Yvette Cedex, France}}\vspace*{0.15cm}
\nopagebreak[3]
\vspace*{-0.20\baselineskip}
\nopagebreak[3]
\address[NIKHEF]{\scriptsize{Nikhef, Science Park,  Amsterdam, The Netherlands}}\vspace*{0.15cm}
\nopagebreak[3]
\vspace*{-0.20\baselineskip}
\nopagebreak[3]
\address[APC]{\scriptsize{APC - Laboratoire AstroParticule et Cosmologie, UMR 7164 (CNRS, Universit\'e Paris 7 Diderot, CEA, Observatoire de Paris) 10, rue Alice Domon et L\'eonie Duquet 75205 Paris Cedex 13,  France}}\vspace*{0.15cm}
\nopagebreak[3]
\vspace*{-0.20\baselineskip}
\nopagebreak[3]
\address[LAM]{\scriptsize{LAM - Laboratoire d'Astrophysique de Marseille, P\^ole de l'\'Etoile Site de Ch\^ateau-Gombert, rue Fr\'ed\'eric Joliot-Curie 38,  13388 Marseille Cedex 13, France }}\vspace*{0.15cm}
\nopagebreak[3]
\vspace*{-0.20\baselineskip}
\nopagebreak[3]
\address[Bologna]{\scriptsize{INFN - Sezione di Bologna, Viale Berti-Pichat 6/2, 40127 Bologna, Italy}}\vspace*{0.15cm}
\nopagebreak[3]
\vspace*{-0.20\baselineskip}
\nopagebreak[3]
\address[Bologna-UNI]{\scriptsize{Dipartimento di Fisica dell'Universit\`a, Viale Berti Pichat 6/2, 40127 Bologna, Italy}}\vspace*{0.15cm}
\nopagebreak[3]
\vspace*{-0.20\baselineskip}
\nopagebreak[3]
\address[Pisa]{\scriptsize{INFN - Sezione di Pisa, Largo B. Pontecorvo 3, 56127 Pisa, Italy}}\vspace*{0.15cm}
\nopagebreak[3]
\vspace*{-0.20\baselineskip}
\nopagebreak[3]
\address[IFIC]{\scriptsize{IFIC - Instituto de F\'isica Corpuscular, Edificios Investigaci\'on de Paterna, CSIC - Universitat de Val\`encia, Apdo. de Correos 22085, 46071 Valencia, Spain}}\vspace*{0.15cm}
\nopagebreak[3]
\vspace*{-0.20\baselineskip}
\nopagebreak[3]
\address[Roma]{\scriptsize{INFN -Sezione di Roma, P.le Aldo Moro 2, 00185 Roma, Italy}}\vspace*{0.15cm}
\nopagebreak[3]
\vspace*{-0.20\baselineskip}
\nopagebreak[3]
\address[Roma-UNI]{\scriptsize{Dipartimento di Fisica dell'Universit\`a La Sapienza, P.le Aldo Moro 2, 00185 Roma, Italy}}\vspace*{0.15cm}
\nopagebreak[3]
\vspace*{-0.20\baselineskip}
\nopagebreak[3]
\address[Clermont-Ferrand]{\scriptsize{Clermont Universit\'e, Universit\'e Blaise Pascal, CNRS/IN2P3, Laboratoire de Physique Corpusculaire, BP 10448, 63000 Clermont-Ferrand, France}}\vspace*{0.15cm}
\nopagebreak[3]
\vspace*{-0.20\baselineskip}
\nopagebreak[3]
\address[GEOAZUR]{\scriptsize{G\'eoazur - Universit\'e de Nice Sophia-Antipolis, CNRS/INSU, IRD, Observatoire de la C\^ote d'Azur and Universit\'e Pierre et Marie Curie, BP 48, 06235 Villefranche-sur-mer, France}}\vspace*{0.15cm}
\nopagebreak[3]
\vspace*{-0.20\baselineskip}
\nopagebreak[3]
\address[Bari]{\scriptsize{INFN - Sezione di Bari, Via E. Orabona 4, 70126 Bari, Italy}}\vspace*{0.15cm}
\nopagebreak[3]
\vspace*{-0.20\baselineskip}
\nopagebreak[3]
\address[COM]{\scriptsize{COM - Centre d'Oc\'eanologie de Marseille, CNRS/INSU et Universit\'e de la M\'editerran\'ee, 163 Avenue de Luminy, Case 901, 13288 Marseille Cedex 9, France}}\vspace*{0.15cm}
\nopagebreak[3]
\vspace*{-0.20\baselineskip}
\nopagebreak[3]
\address[LNS]{\scriptsize{INFN - Laboratori Nazionali del Sud (LNS), Via S. Sofia 62, 95123 Catania, Italy}}\vspace*{0.15cm}
\nopagebreak[3]
\vspace*{-0.20\baselineskip}
\nopagebreak[3]
\address[UPS]{\scriptsize{Univ Paris-Sud , 91405 Orsay Cedex, France}}\vspace*{0.15cm}
\nopagebreak[3]
\vspace*{-0.20\baselineskip}
\nopagebreak[3]
\address[KVI]{\scriptsize{Kernfysisch Versneller Instituut (KVI), University of Groningen, Zernikelaan 25, 9747 AA Groningen, The Netherlands}}\vspace*{0.15cm}
\nopagebreak[3]
\vspace*{-0.20\baselineskip}
\nopagebreak[3]
\address[Pisa-UNI]{\scriptsize{Dipartimento di Fisica dell'Universit\`a, Largo B. Pontecorvo 3, 56127 Pisa, Italy}}\vspace*{0.15cm}
\nopagebreak[3]
\vspace*{-0.20\baselineskip}
\nopagebreak[3]
\address[NIOZ]{\scriptsize{Royal Netherlands Institute for Sea Research (NIOZ), Landsdiep 4,1797 SZ 't Horntje (Texel), The Netherlands}}\vspace*{0.15cm}
\nopagebreak[3]
\vspace*{-0.20\baselineskip}
\nopagebreak[3]
\address[Bamberg]{\scriptsize{Dr. Remeis-Sternwarte and ECAP, Universit\"at Erlangen-N\"urnberg,  Sternwartstr. 7, 96049 Bamberg, Germany}}\vspace*{0.15cm}
\nopagebreak[3]
\vspace*{-0.20\baselineskip}
\nopagebreak[3]
\address[UU]{\scriptsize{Universiteit Utrecht, Faculteit Betawetenschappen, Princetonplein 5, 3584 CC Utrecht, The Netherlands}}\vspace*{0.15cm}
\nopagebreak[3]
\vspace*{-0.20\baselineskip}
\nopagebreak[3]
\address[UvA]{\scriptsize{Universiteit van Amsterdam, Instituut voor Hoge-Energie Fysika, Science Park 105, 1098 XG Amsterdam, The Netherlands}}\vspace*{0.15cm}
\nopagebreak[3]
\vspace*{-0.20\baselineskip}
\nopagebreak[3]
\address[MSU]{\scriptsize{Moscow State University,Skobeltsyn Institute of Nuclear Physics,Leninskie gory, 119991 Moscow, Russia}}\vspace*{0.15cm}
\nopagebreak[3]
\vspace*{-0.20\baselineskip}
\nopagebreak[3]
\address[Catania]{\scriptsize{INFN - Sezione di Catania, Viale Andrea Doria 6, 95125 Catania, Italy}}\vspace*{0.15cm}
\nopagebreak[3]
\vspace*{-0.20\baselineskip}
\nopagebreak[3]
\address[Catania-UNI]{\scriptsize{Dipartimento di Fisica ed Astronomia dell'Universit\`a, Viale Andrea Doria 6, 95125 Catania, Italy}}\vspace*{0.15cm}
\nopagebreak[3]
\vspace*{-0.20\baselineskip}
\nopagebreak[3]
\address[IRFU/SPP]{\scriptsize{Direction des Sciences de la Mati\`ere - Institut de recherche sur les lois fondamentales de l'Univers - Service de Physique des Particules, CEA Saclay, 91191 Gif-sur-Yvette Cedex, France}}\vspace*{0.15cm}
\nopagebreak[3]
\vspace*{-0.20\baselineskip}
\nopagebreak[3]
\address[Sheffield]{\scriptsize{Department of Physics and Astronomy, University of Sheffield, Hicks Building, Hounsfield Road, Sheffield S3 7RH, UK}}\vspace*{0.15cm}
\nopagebreak[3]
\vspace*{-0.20\baselineskip}
\nopagebreak[3]
\address[WIN]{\scriptsize{DPNC, Universit\'e de Gen\`eve, Quai Ernest-Ansermet 24, 1211 Gen\`eve, Switzerland}}\vspace*{0.15cm}
\nopagebreak[3]
\vspace*{-0.20\baselineskip}
\nopagebreak[3]
\address[ISS]{\scriptsize{Institute for Space Sciences, R-77125 Bucharest, M\u{a}gurele, Romania     }}\vspace*{0.15cm}
\nopagebreak[3]
\vspace*{-0.20\baselineskip}
\nopagebreak[3]
\address[IPHC]{\scriptsize{IPHC-Institut Pluridisciplinaire Hubert Curien - Universit\'e de Strasbourg et CNRS/IN2P3  23 rue du Loess, BP 28,  67037 Strasbourg Cedex 2, France}}\vspace*{0.15cm}
\nopagebreak[3]
\vspace*{-0.20\baselineskip}
\nopagebreak[3]
\address[ITEP]{\scriptsize{ITEP - Institute for Theoretical and Experimental Physics, B. Cheremushkinskaya 25, 117218 Moscow, Russia}}\vspace*{0.15cm}
\nopagebreak[3]
\vspace*{-0.20\baselineskip}
\nopagebreak[3]
\address[Genova-UNI]{\scriptsize{Dipartimento di Fisica dell'Universit\`a, Via Dodecaneso 33, 16146 Genova, Italy}}\vspace*{0.15cm}
\nopagebreak[3]
\vspace*{-0.20\baselineskip}

\begin{abstract}
The group velocity of light has been measured at eight different
wavelengths between 385~nm and 532~nm in the Mediterranean Sea at a
depth of about 2.2~km with the ANTARES optical beacon systems. 
A parametrisation of the dependence of the refractive index 
on wavelength based on the salinity, pressure and temperature 
of the sea water at the ANTARES site is in good agreement 
with these measurements.\end{abstract}


\end{frontmatter}

\newpage


\section{Introduction}
\label{sec:introduction}
The ANTARES neutrino telescope is located on the bottom of the
Mediterranean Sea ($42^o 50'$ N, $6^o 10'$ E) at a depth of 2475~m, roughly \mbox{40~km} offshore
from Toulon in France. The main objective of the experiment is the observation
of neutrinos of cosmic origin in the southern hemisphere sky.
Sea water is used as the detection medium of
the Cherenkov light induced by relativistic charged particles resulting
from the interaction of neutrinos.  The particle
trajectory is reconstructed from the measured arrival times of the
detected photons. The detector consists of 885
photomultiplier tubes (PMTs) mounted on twelve vertical lines with
a length of about 450~m. The horizontal separation
between lines is about 70~m.  Further details can be found elsewhere~\cite{Ant1,Ant2,DAQ}.

Charged particles traveling through sea water produce the emission of Cherenkov
light whenever the velocity of the particle exceeds that of light in water. 
The Cherenkov photons are emitted at a characteristic angle, $\theta_c$, 
with respect
to the particle direction. This angle is related to the index
of refraction of the medium as $cos \, \theta_c = \frac{1}{\beta n_p}$. 
In this, $\beta$ is the velocity of the particle relative 
to the speed of light in vacuum.
The index of refraction, $n_p$, corresponds to the ratio between the speed 
of light in vacuum and the phase velocity of light in water.
The individual photons then travel through the water at the group velocity.
Both the phase and the group velocity depend on the
wavelength of the photons. This is usually referred to as 
chromatic dispersion.  The group
velocity is related to its phase velocity in the following way:
\begin{equation}
n_g=\frac{n_p}{1+\frac{\lambda}{n_p}\frac{d n_p}{d \lambda}}
\label{equ:n_g}
\end{equation}
\noindent where $\lambda$ is the wavelength of light. 
The index of refraction, $n_g$, corresponds to the ratio between 
the speed of light in vacuum and the group velocity of light in water.

Since the PMTs cannot distinguish the photon wavelength, 
the variation of the photon emission
angle and the group velocity due to chromatic dispersion cannot be
accounted for on the individual photon level. Nevertheless, 
the average effect of the wavelength dependencies are accounted for in 
the algorithm used to reconstruct the particle trajectory~\cite{Heijboer,Coll2011}.

A measurement of the group velocity of light has been made using
the optical beacon system of ANTARES. This system consists of 
a set of pulsed light sources
(LEDs and lasers) which are distributed throughout the detector and illuminate the
PMTs with short duration flashes of light.  The refractive index is
deduced from the recorded time of flight distributions of photons at
different distances from the sources for eight
different wavelengths between 385~nm and 532~nm.

\section{Experimental Setup}
\label{sec:expsetup}
The PMTs of ANTARES are sensitive to photons in the wavelength range
between 300~nm and 600~nm. The maximum quantum efficiency is
about 22~\% between 350~nm and 450~nm. The arrival
time and integrated charge of the analogue pulse from the PMT are measured
by the readout electronics~\cite{AguiFront}.  
The transit time spread of single photo-electrons of the PMT is around 3.5~ns
(FWHM)~\cite{Timecalibration}.

The group velocity of light has been measured using the ANTARES optical
beacon system. This system was primarily
designed to perform time calibration {\it in situ}~\cite{Timecalibration,Ageron}. 
There are two types of optical beacons, the
LED optical beacons and the laser beacons. There are four LED optical beacons
distributed along each detector line and two laser beacons at the bottom of two central lines.
The {\it in situ} measurement of the temperature and salinity 
is provided by some Conductivity/Temperature/Depth sensors\footnote{SEABIRD CTD (SBE37-SMP), http://www.seabird.com/.}. 

\begin{figure}
 \begin{center}
 \begin{tabular}{c c c}
 \epsfysize=3.4cm
 \epsffile{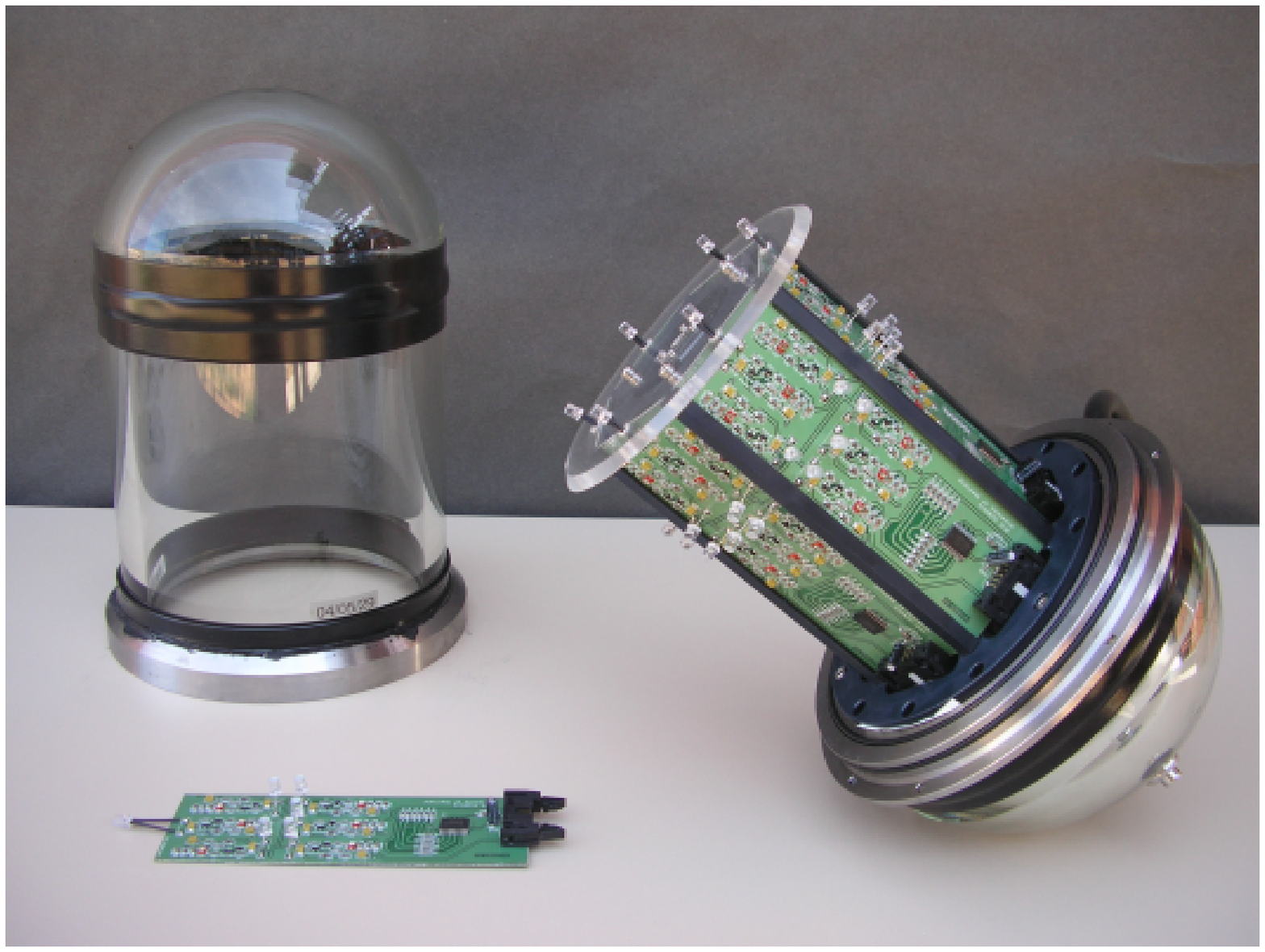}
 \epsfysize=3.4cm
 \epsffile{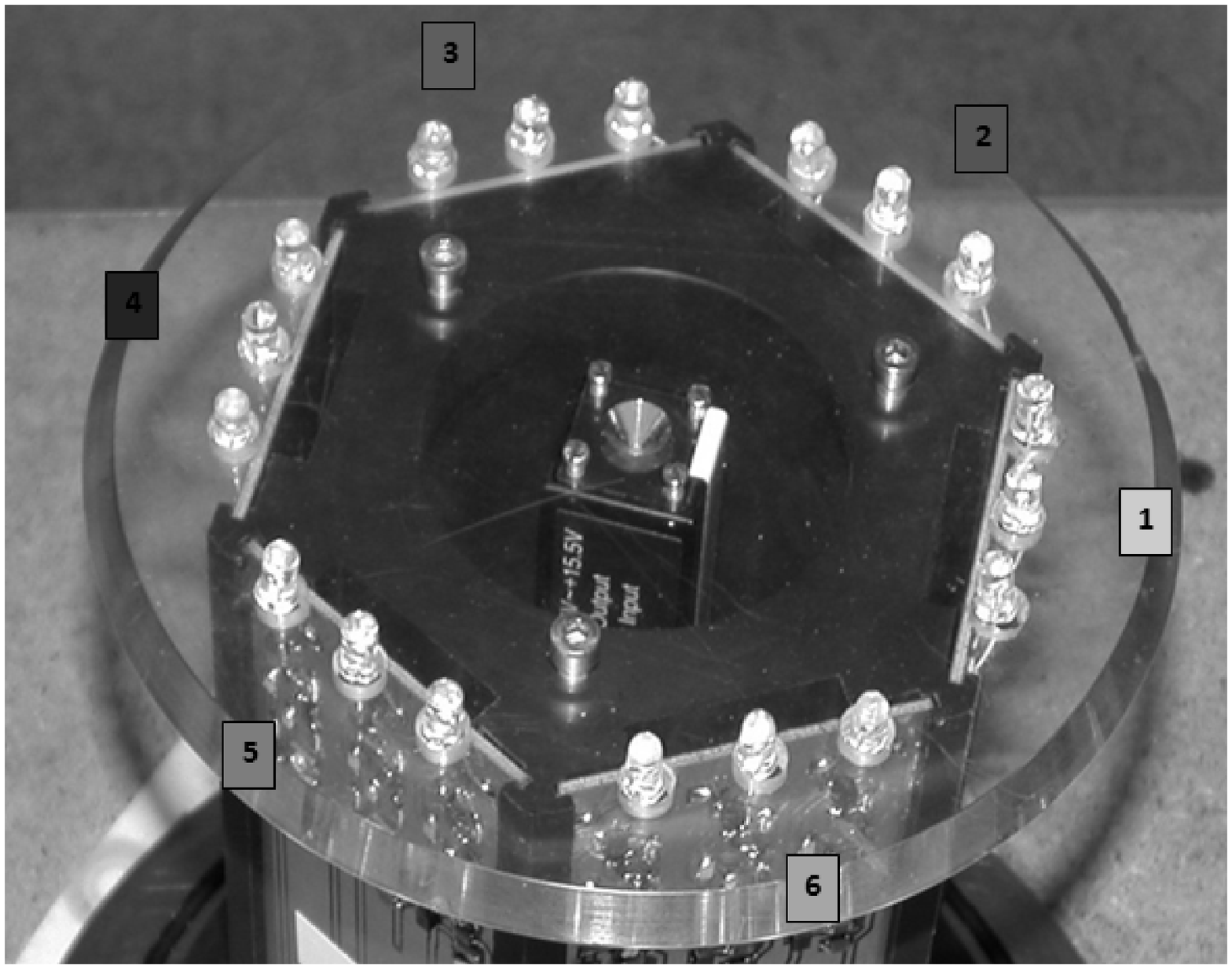}
 \epsfysize=3.4cm
 \epsffile{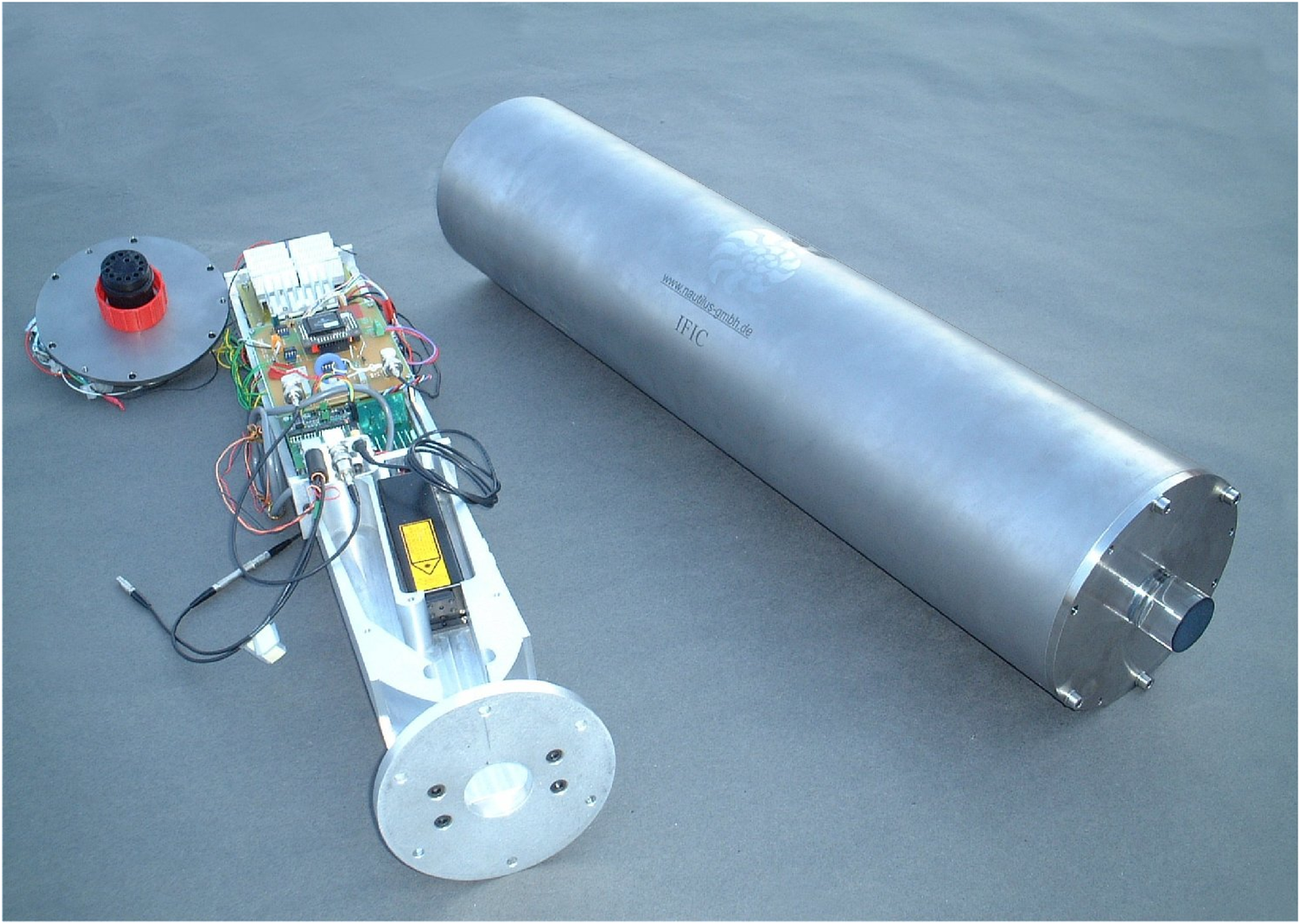}
 \end{tabular}
 \caption{Picture of a standard LED optical beacon (left), the modified LED optical beacon (middle) and a laser beacon (right).}
 \label{fig:opticalb}
 \end{center}
\end{figure}

A standard LED optical beacon contains 36 individual LEDs distributed over six vertical faces
forming a hexagonal cylinder housed in a pressure resistant glass enclosure (Figure~\ref{fig:opticalb}, left). On
each face, five LEDs point radially outwards and one upwards. All 
LEDs emit light at an average wavelength of 469~nm, except the two LEDs
located on the lowest LED optical beacon of Line 12 which emit light at an average
wavelength of 400~nm. A modified LED optical beacon was installed in
2010 on Line 6.  This LED optical beacon has three LEDs per face instead of six, all of them pointing
upwards (Figure~\ref{fig:opticalb}, centre).  The three LEDs of each
face emit light of the same colour.
The average wavelength of the light from the six faces are 385, 400, 447, 458, 494 and 518~nm.
The LEDs emit 
light with a maximum intensity of  about 160~pJ and a
pulse width of about 4~ns FWHM. The intensity of the emitted light
can be varied changing the voltage feeding the LEDs.

\label{detector}

The laser beacon consists of a Nd-YAG solid state laser (Figure~\ref{fig:opticalb}, right). It can 
emit pulses of light with an intensity of  about 1~$\mu$J and pulse
width of about 0.8~ns FWHM. The average wavelength is 532~nm.
This light is spread out by an optical diffuser, so that the light
can reach the surrounding lines.
During calibration runs, the LEDs and lasers flash at a
frequency of 330~Hz.  Further
details about the optical beacon system can be found elsewhere~\cite{Timecalibration,Ageron,Juanan}.

\begin{figure}
 \setlength{\unitlength}{1cm}
 \centering
 \begin{picture}(18.5,7.5)
   \put(-1.0,3.5){\epsfig{file=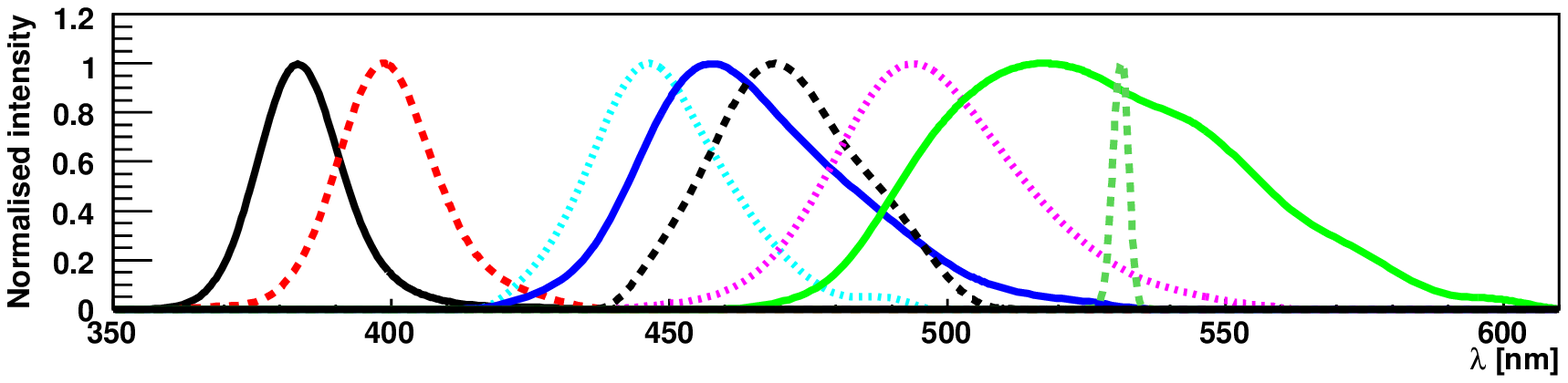,width=17.0cm,clip=}}
   \put(-1.0,-0.3){\epsfig{file=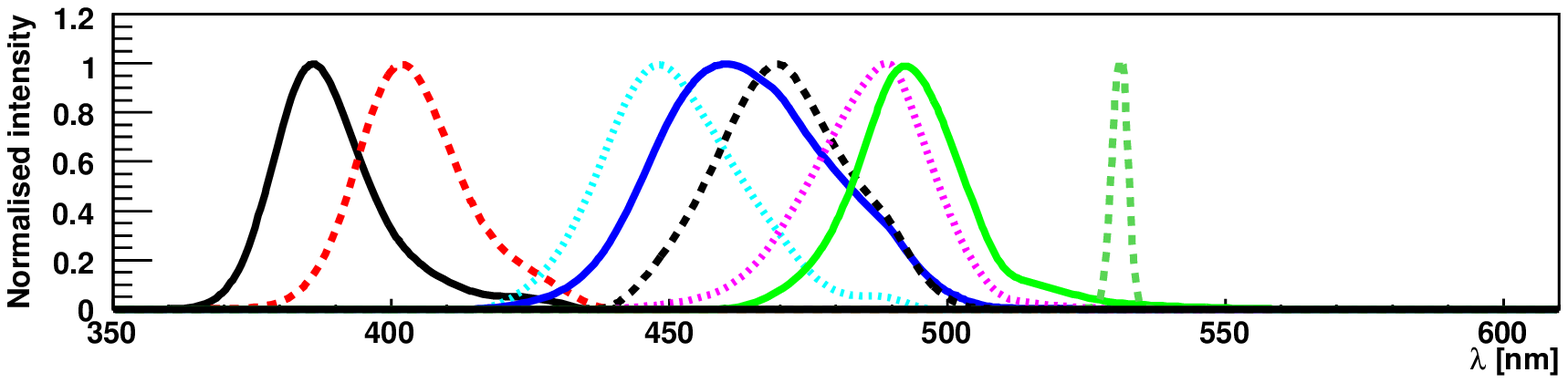,width=17.0cm,clip=}}
   \put(0.7,6.6){\textbf{\textsf{(a)}}}
   \put(0.7,2.8){\textbf{\textsf{(b)}}}
   \put(1.7,7.1){\scriptsize{\textbf{\textsf{385}}}}
   \put(2.5,7.1){\scriptsize{\textbf{\textsf{400}}}}
   \put(5.0,7.1){\scriptsize{\textbf{\textsf{447}}}}
   \put(5.6,7.1){\scriptsize{\textbf{\textsf{458}}}}
   \put(6.25,7.1){\scriptsize{\textbf{\textsf{469}}}}
   \put(7.5,7.1){\scriptsize{\textbf{\textsf{494}}}}
   \put(8.6,7.1){\scriptsize{\textbf{\textsf{518}}}}
   \put(9.5,7.1){\scriptsize{\textbf{\textsf{532}}}}
 \end{picture}
\caption[Sc]{(a) Wavelength spectra of the light sources used in this work as measured with a calibrated spectrometer. The spectrometer measurements are made in air. The data points have been smoothed and the highest value of each spectrum is set equal to one. On the top of the Figure the peak wavelengths are indicated in units of nm. (b) Simulated light spectra at a distance of 120~m in sea water with the highest value renormalized to one. The differences between the spectra are due to the variation of the absorption length as a function of the wavelength.}
\label{fig:specmeas}
\end{figure}

The wavelength spectra of the light sources used for this analysis were measured using a calibrated
spectrometer\footnote{Ocean Optics HR4000CG-UV-NIR, http://www.oceanoptics.com/.}(see Figure~\ref{fig:specmeas}a). 
The typical width of each spectrum
is around 10~nm except for the green LED (518~nm), which is larger, and the laser (532~nm), which is
much smaller. 

Due to the wavelength dependence of the absorption of light in water, 
the spectra change as a function of the distance travelled by the light.
The expected wavelength distributions as a
function of distance have been estimated by Monte Carlo simulations
using the dependence of absorption length on wavelength given by Smith
and Baker~\cite{abs}. In Figure~\ref{fig:specmeas}b the spectra at a distance of 
120~m are shown for the different light sources.
In particular, absorption has a large effect for wavelengths above 500~nm. Notice that 
the distributions have been renormalised to unity in each peak and 
therefore the relative effect of absorption between sources is not observed.  
This renormalisation is performed in order to show 
the change in the shapes of the distributions, which is what influences the velocity measurement.
The evolution of the spectra is taken into account in the final results (Section~\ref{sec:data}),
in particular the uncertainty assigned to the wavelengths has been taken to be the root mean 
square (RMS) of the wavelength distribution given by the simulation.

\section{Data Acquisition and Analysis}
\label{sec:dataacqu}
In order to measure the optical properties of the deep sea water, designated data taking 
runs were performed using the optical beacon system. During these
runs, one single LED located in the lowest optical beacon of a line and pointing upward 
was flashed. Only the signals recorded by the
PMTs along the same line are used in
the analysis. As a result, the line movements due to the sea currents can safely be ignored.

The runs used in this analysis were taken between May 2008 and April 2011.
Each run contains typically more than 100,000 light flashes. Each flash is detected by a 
small PMT inside the optical beacon. The time of the flash and the arrival times of the photons 
on the PMTs were recorded within a time window from 1500~ns
before to 1500~ns after the flash. The integrated charge of the analogue pulses of the PMTs were also recorded. 
Only runs were used when the average rate of background light was below 100~kHz.

\begin{figure}
 \setlength{\unitlength}{1cm}
 \centering
 \begin{picture}(18.5,6.5)
   \put(0.5,0.0){\epsfig{file=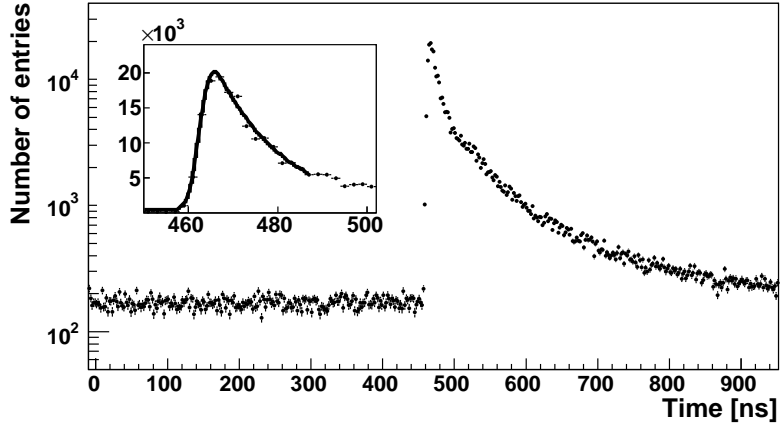,width=12.0cm,clip=}}
 \end{picture}
\caption[Sc]{Example of a distribution of arrival times observed at a distance of 100~m from an LED. The inset shows a zoom around the signal region. The solid curve corresponds to a fit function as described in the text and given in Equation \ref{equ:ciro}.}
\label{fig:time}
\end{figure}

In Figure \ref{fig:time}, the distribution of the arrival times of
photons on a PMT located 100~m above the LED optical beacon ($\lambda$ = 469~nm) is shown. 
The time, \mbox{$t$ = 0~ns}, corresponds to the time of the flash.
A clear peak at $t$ = 470~ns can be seen which corresponds to the shortest propagation time of the light.
The tail with late photons can be attributed to light scattering.  The flat background arises from 
the optical background due to $^{40}$K decays and
bioluminescence.

A convolution of a Gaussian and an exponential distribution on top of a flat background
is fitted to the data.
The Gaussian distribution reproduces the
transit time spread of the PMTs, the duration of the light flash and the effect of the chromatic dispersion
in water. The exponential distribution takes into account the effect of the scattering of
photons in water.  The fit function can be formulated as:
\begin{equation}
  f(t)=B + S \cdot e^{-\frac{\displaystyle t-\mu}{\displaystyle \tau}} \times \textrm{erfc} \Bigg(\frac{1}{\sqrt{2}} \Big(\frac{\sigma}{\tau} - \frac{t-\mu}{\sigma}\Big)\Bigg)
\label{equ:ciro}
\end{equation}
where $t$ is the arrival time of the photons. The fit parameters
are the optical background, $B$, the signal strength, $S$, 
the mean, $\mu$,  and width, $\sigma$, of the Gaussian distribution
and the exponential decay constant, $\tau$. In Equation \ref{equ:ciro}, $\textrm{erfc}(t)$ is the complementary error function
distribution.  An example of the fit is shown in the inset of
Figure~\ref{fig:time}. The
fit is determined in the range from 200~ns before the most populated bin and
20~ns after.  The arrival time of the light flash at each PMT is estimated by the fitted mean
value of the Gaussian distribution.

\begin{figure}[t!]
 \setlength{\unitlength}{1cm}
 \centering
 \begin{picture}(18.5,8.5)
  \put(-1.3,0.3){\epsfig{file=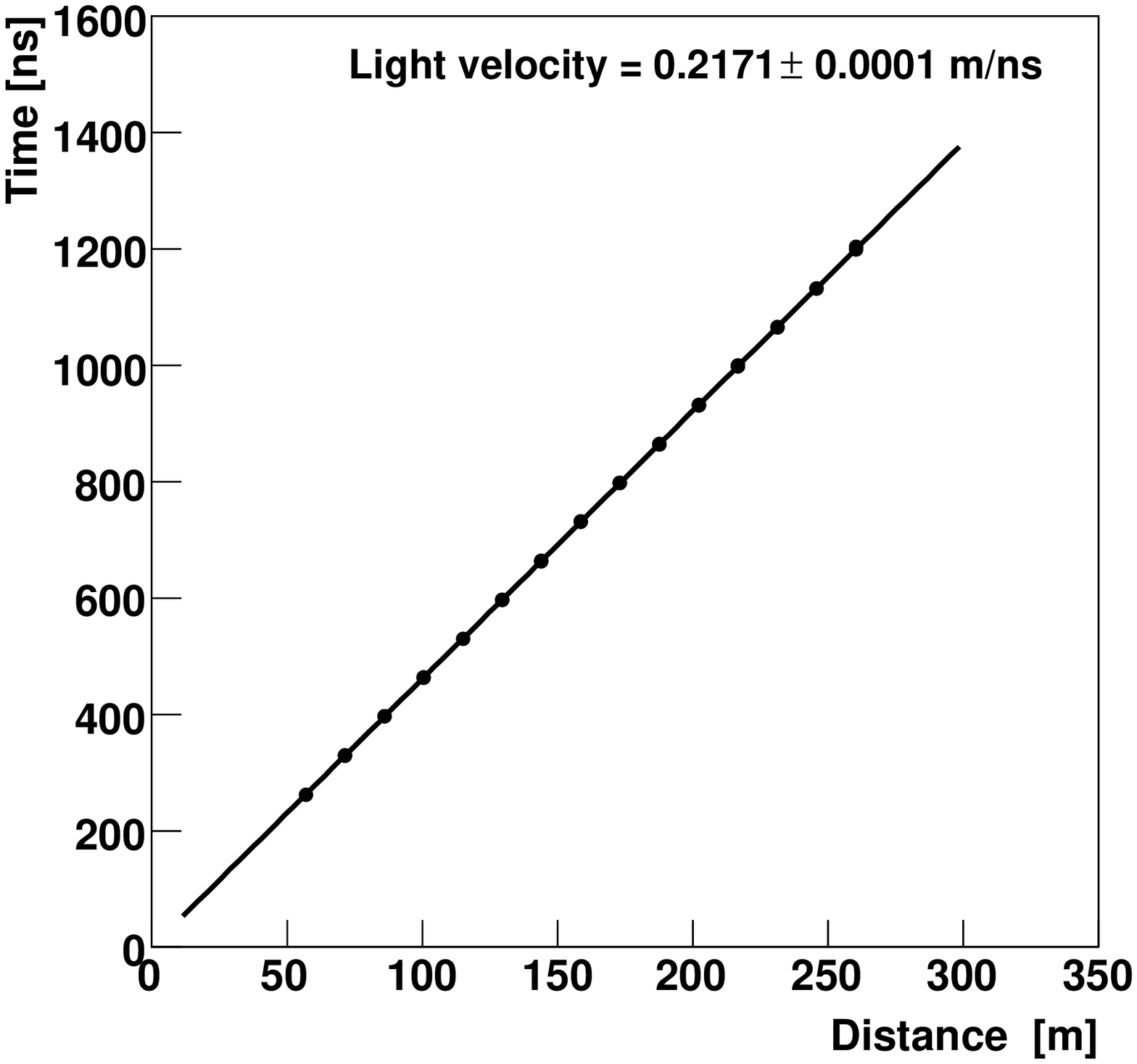,width=8.0cm,clip=}}
  \put(6.5,0.3){\epsfig{file=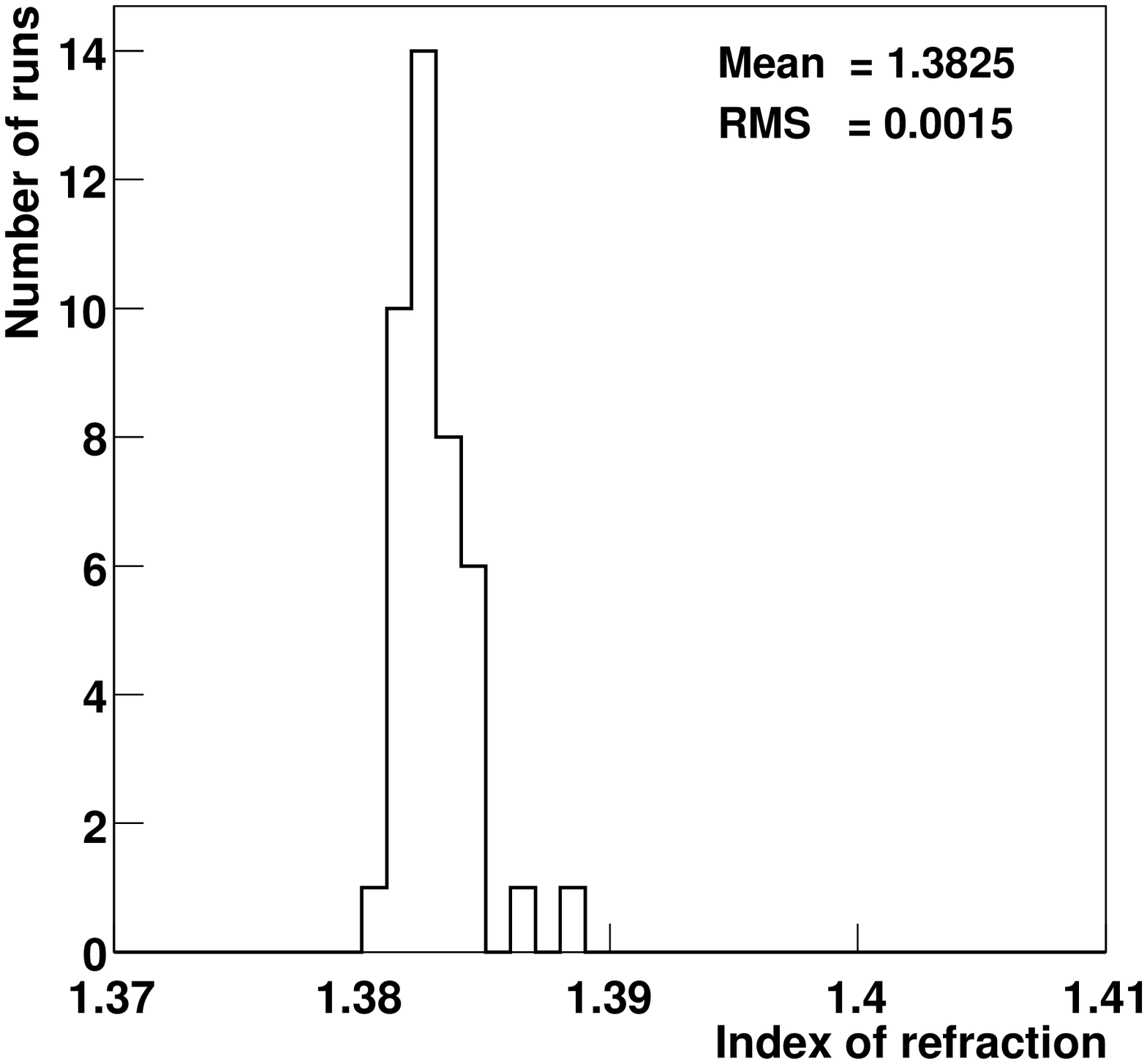,width=8.0cm,clip=}}
  \put(0.1,6.7){\textbf{\textsf{(a)}}}
  \put(7.8,6.7){\textbf{\textsf{(b)}}}
 \end{picture}
\caption[Sc]{(a) Arrival time as a function of the distance between the LED ($\lambda$ = 469~nm) and the PMT for one single run. The solid line corresponds to a fit of a linear function to the data (see text). (b) Distribution of the measured refractive index for a total of 42 runs.}
\label{fig:linearfit}
\end{figure}

An example of the measured arrival times as a function of the distance between the
optical beacon and the PMT is shown in
Figure~\ref{fig:linearfit}a. 
A linear function has been fitted to the data to extract the group velocity of the light. 
In the fitting procedure, the distance between the light source 
and the PMT has been restricted to the 
range between 50~m and 250~m in order to limit the effects of time slewing~\cite{Timecalibration} and optical background on the measurement.

A Monte Carlo simulation of the response of the detector to LED flashes has been made.
The analysis method was 
performed to validate the analysis procedure and to study the systematic effects
due to the assumed light absorption and scattering parameters.
Table \ref{tab:Systall} shows the different contributions to the systematic 
uncertainty in the measurement of the refractive index. 
These contributions have been determined as follows. For the wavelength values of 400~nm, 470~nm and 532~nm, 
the difference of the refractive index obtained after the fitting procedure with respect to the 
input refractive index is shown in the first row of Table \ref{tab:Systall}. 
This variation, which has been computed for the 
default values of absorption and scattering lengths, is termed "Method". In the second row, the 
variation in percentage of the fitted value is shown 
when the absorption length changes by $\pm$50~\%. Finally, in the third 
row the variation in percentage of the fitted value is shown when the scattering length is varied 
between 20~m and 70~m~\cite{Agui}. As can be seen, the latter is the largest contribution to the systematic 
uncertainty. Adding in quadrature these values one obtains a systematic error that 
varies from 0.25~\% to 0.37~\% depending on the wavelength.

\begin{table}[tcb]
\begin{center}
\begin{tabular}{|c|r|r|r|}
\hline
Wavelength [nm] & 400  & 470 & 530  \\
\hline
Source of uncertainty & \multicolumn{3}{|c|}{Variation in [\%]} \\
\hline
Method & 0.27 &  0.13 & 0.04  \\
\hline
Absorption length& 0.11 & 0.05&  0.07 \\
\hline
Scattering length & 0.22& 0.34&  0.24 \\
\hline
Total systematic uncertainty [\%] & $\pm$0.37 &$\pm$0.37 & $\pm$0.25  \\
\hline
\end{tabular}
\end{center}
\vspace{10mm}
\caption[]{\textit{The systematic uncertainty of the refractive index measurement is estimated for three different wavelengths by varying the absorption length and the scattering length (see text).}}
\label{tab:Systall}
\end{table}

\section{Determination of the Refractive Index}
\label{sec:data}
\begin{figure}[t]
 \setlength{\unitlength}{1cm}
 \centering
 \begin{picture}(18.5,9.5)
  \put(-0.5,0.3){\epsfig{file=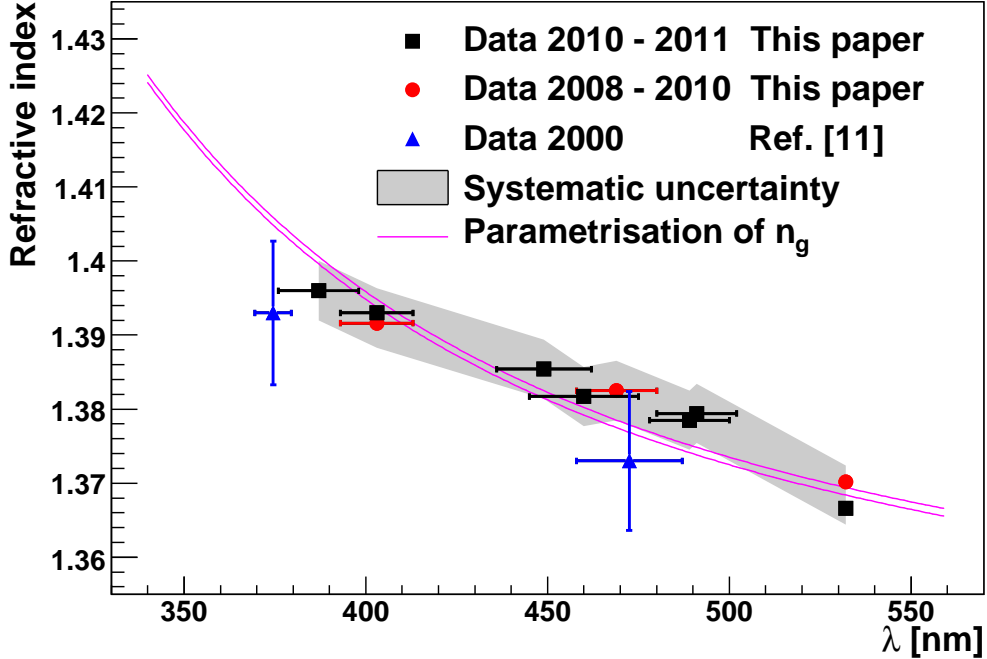,width=14.5cm,clip=}}
 \end{picture}
\caption[Sc]{Index of refraction corresponding to the group velocity of light as a function of the wavelength. Also shown are results from measurements made in~\cite{Agui}. The grey band shows the systematic uncertainty. The two solid lines correspond to a parametrisation of the index of refraction evaluated at a pressure of 200 atm (lower line) and 240 atm (upper line) (see text).}
\label{fig:refdat}
\end{figure}

Between May 2008 and March 2010, a total of 42 runs were taken  using an LED with 
an average wavelength of 469~nm. Three different
LED intensities were used. For a high, middle or low
intensity run the range of distances between the optical beacon and 
the PMT used in the fit were 50~--~250~m, 40~--~220~m and 10~--~130~m, respectively.
The measured refractive index values of these runs are shown
in Figure~\ref{fig:linearfit}b. In addition to these runs, 14 runs using an LED with 
an average wavelength of 400~nm and 13 runs using an LED with an average wavelength of 532~nm were taken. 
Between November 2010 and April 2011 eight runs with a modified 
optical beacon were taken, extending the measurements with six additional wavelengths.
The index of refraction is estimated at each wavelength by the mean of the distribution.
The measured refractive indices with the systematic uncertainty are shown in Figure~\ref{fig:refdat}
and tabulated in Table~\ref{table}. As mentioned in Section~\ref{sec:expsetup}, the uncertainties in the wavelengths
have been taken to be the RMS of the corresponding distribution at the middle of the distance ranges. The 
variation of the RMS values in this range with respect to the middle is $\pm$2~nm.  

\begin{table}[]
\centering
\begin{tabular}{|c| c| c| c|}
\hline
Wavelength (nm)  & Refractive index & Number of runs & Time period\\
\hline
403 $\pm$ 10      & 1.3916 $\pm$ 0.0007 &  14 &2008-2010\\
469 $\pm$ 12      & 1.3825 $\pm$ 0.0002 &  42 &2008-2010\\
532 $\pm$ ~~1     & 1.3702 $\pm$ 0.0007 &  13 &2008-2010\\
\hline
387 $\pm$ 11      & 1.3960 $\pm$ 0.0007 &  8  &2010-2011\\
403 $\pm$ 10      & 1.3930 $\pm$ 0.0007 &  8  &2010-2011\\
449 $\pm$ 13      & 1.3854 $\pm$ 0.0003 &  8  &2010-2011\\
460 $\pm$ 15      & 1.3817 $\pm$ 0.0003 &  8  &2010-2011\\
489 $\pm$ 11      & 1.3785 $\pm$ 0.0003 &  8  &2010-2011\\
491 $\pm$ 11      & 1.3794 $\pm$ 0.0006 &  8  &2010-2011\\
532 $\pm$ ~~1     & 1.3666 $\pm$ 0.0006 &  8  &2010-2011\\
\hline
\end{tabular}
\vspace*{1.8\baselineskip}
\caption[Sc]{Summary of the refractive index results for the 2008 - 2011 data shown in \mbox{Figure~\ref{fig:refdat}}. For the refractive index first the statistical uncertainties are shown. They are computed as the RMS of the measured refractive index for each wavelength divided by the square root of the number of runs. In addition, there is a total systematic uncertainty of $\pm$ 0.005.}
\label{table}
\end{table}

The velocity of light in sea water at a given wavelength depends
on the temperature, the salinity and the pressure of the water, because
the density of sea water depends on these variables.
A parametrisation of the light velocity 
proposed by \mbox{Quan and Fry}~\cite{Quanfry} is based
on data from Austin and Halikas~\cite{Austin}. This parametrisation was modified 
to incorporate a correction for pressure~\cite{Agui}. 
During the data taking period, the temperature and salinity were measured in situ at a depth of 2250 m. 
At an ambient temperature of 
\mbox{T = $12.9 \pm 0.1~^o$C} and salinity of S = $38.48 \pm 0.01~\permil$, the refractive
index corresponding to the phase velocity as a function of wavelength is expressed as:
\begin{multline}
n_p (\lambda, p)= 1.32292+(1.32394-1.32292) \times \frac{p-200}{240-200}+ \\
+\frac{16.2561}{\lambda}-\frac{4382}{\lambda^2}+\frac{1.1455\times10^6}{\lambda^3}
\label{equ:nphase}
\end{multline}
where $\lambda$ is the wavelength (in units of nm) and $p$ is the pressure (in units of atm).
Using Equation \ref{equ:n_g}, the result of this parametrisation can be compared to the measurements 
(see \mbox{Figure~\ref{fig:refdat}}). 

From the known variations of temperature, salinity and pressure, the
refractive index for a particular wavelength and at a given depth can
be determined with an accuracy of better than $4 \times 10^{-5}$.
The parametrisation is in good agreement with the measurements.

As mentioned in Section~\ref{sec:introduction}, the PMTs are unable to distinguish the wavelength
of the incoming photons, so the effect of this chromatic dependence
can only be taken into account on average. 
The spread of the arrival time residuals with
respect to the expected arrival time of a 460~nm photon have been
computed by means of a standalone Monte Carlo simulation using the
phase velocity for the emission angle and the group velocity (as given
by the Equations \ref{equ:nphase} and \ref{equ:n_g}) for the arrival time.
This simulation indicates that the spread of the time residual is
0.6~ns at 10~m, 1.6~ns at 40~m, 2.7~ns at 100~m and 3.6~ns at 200~m. The
time uncertainty introduced by this spread is unavoidable and is taken
into account in the ANTARES official simulation program~\cite{KM3,Bru}.
Even though the exact influence of the medium depends on the particular Cherenkov
photon (wavelength, distance to the hit PMT) and therefore requires a
full simulation, a rough estimate of the average effect can be
obtained assuming that a majority of hits are between 40~m and 100~m
from the track, which gives a value of $\sim$2~ns for the uncertainty
introduced by the transmission of light in sea water, including
chromatic dispersion. This value is to be compared with $\sim$1.3~ns
coming from the PMTs transit time spread and to $\sim$1~ns from time
calibration. Monte Carlo simulation studies show also that an additional
uncertainty of $\sim$2~ns on the spread of the time residual degrades the
neutrino pointing accuracy about 0.1 degrees.

\section {Summary}

Using pulsed light sources with wavelengths between 385~nm and 532~nm
the group velocity of light in sea water at the ANTARES site has been
measured as a function of wavelength. The emission spectra determined
in the laboratory for the different pulsed sources have been used as
input to a Monte Carlo simulation in order to correct for the effect
of absorption on the corresponding velocity measurement. Except for
two sources these corrections are in
general small. Likewise, a Monte Carlo simulation has been used to
evaluate the systematic uncertainties and to
check, that the procedure to obtain the speed of light is robust and
unbiased. 
The results obtained
for the dependence of the group refractive index on wavelength are in
agreement with the parametrisation as a function of salinity, pressure
and temperature of sea water at the ANTARES site.

\section*{Acknowledgments}

The authors acknowledge the financial support of the funding agencies:
Centre National de la Recherche Scientifique (CNRS), Commissariat
\'a l'\'ene\-gie atomique et aux \'energies alternatives  (CEA), Agence
National de la Recherche (ANR), Commission Europ\'eenne (FEDER fund
and Marie Curie Program), R\'egion Alsace (contrat CPER), R\'egion
Provence-Alpes-C\^ote d'Azur, D\'e\-par\-tement du Var and Ville de
La Seyne-sur-Mer, France; Bundesministerium f\"ur Bildung und Forschung
(BMBF), Germany; Istituto Nazionale di Fisica Nucleare (INFN), Italy;
Stichting voor Fundamenteel Onderzoek der Materie (FOM), Nederlandse
organisatie voor Wetenschappelijk Onderzoek (NWO), the Netherlands;
Council of the President of the Russian Federation for young scientists
and leading scientific schools supporting grants and Rosatom, Russia; National
Authority for Scientific Research (ANCS), Romania; Ministerio de Ciencia
e Innovaci\'on (MICINN), Prometeo of Generalitat Valenciana and MultiDark,
Spain. We also acknowledge the technical support of Ifremer, AIM and
Foselev Marine for the sea operation and the CC-IN2P3 for the computing facilities.

\bibliographystyle{report}  




\bibliography{sabib}







\end{document}